\documentclass[prb,aps,preprint,eqsecnum]{revtex4}
\usepackage{epsfig}
\usepackage{amssymb}

\begin{document}

\title{A generalization of Wolynes factor in activated processes}
\author{Debashis~Barik and Deb~Shankar~Ray{\footnote{Email Address:
pcdsr@mahendra.iacs.res.in}}} \affiliation{Indian Association for
the Cultivation of Science, Jadavpur, Kolkata 700 032, India.}

\begin{abstract}
Kramers-Grote-Hynes factor is the key nonequilibrium contribution
to rate constant of a reaction over and above the transition
state theory rate in the spatial limited regime. Wolynes in
eighties introduced a quantum correction to the overall rate
coefficient. This is responsible for tunneling and quantum
enhancement of rate at low temperature. However, its validity is
restricted to activated tunneling region or above crossover
temperature. Based on a quantum formulation of the normal mode
analysis, we suggest a generalization of Wolynes factor and a
consequent multidimensional transition state rate expression
which are valid in the deep tunneling region down to zero degree
Kelvin.
\end{abstract}

\maketitle

\section{{\bf Introduction}}
Kramers' \cite{k1} diffusion model of chemical reactions proposed
in 1940 forms the dynamical basis of modern rate theory of
activated processes. The seminal and essential content of this
nonequilibrium formulation is the inclusion of dependence of rate
constant of a reaction on viscosity or friction of the reaction
medium. Based on the classical theory of Brownian motion in phase
space Kramers derived the expressions for nonequilibrium steady
state distribution functions to work out the rate coefficients in
the two different limiting situations and showed that the rate
varies linearly in the weak dissipation regime and inversely in
the high dissipation regime with friction. With the advent of
ultrafast lasers and time-resolved detection techniques since
late seventies, experimental confirmation of Kramers' theory
provided a new impetus for further development in chemical
dynamics and condensed matter physics \cite{dn,cs,lut,man,ot}.

While Kramers' theory is based on classical Markovian
description, eighties saw a number of important generalization of
Kramers theory in several directions by a number of groups
\cite{gr,car,uw,str,mel,h1,san,tan,co}. An important endeavor in
this context is the inclusion of a quantum equilibrium factor,
known as Wolynes factor \cite{woly,wei}, in the generalized
Kramers' or equivalently multidimensional transition state theory
rate constant. This term allows tunneling as a possible decay
route of the metastable state in addition to the usual thermally
activated processes and is responsible for quantum enhancement of
rate at low temperature. However, the validity of the Wolynes
factor is restricted to the activated tunneling regime or above
crossover temperature. Based on a positive definite Wigner
function formalism within a normal mode description we generalize
Wolynes factor and thereby multidimensional transition state rate
expression which are valid in the deep tunneling region and
vacuum limit.

The outlay of the paper is as follows. In the following Sec.II we
discuss a quantum Langevin equation which allows us to realize a
c-number Hamiltonian amenable to a normal mode analysis. The
equilibrium theory in terms of Wigner distribution function to
formulate a quantum counterpart of multidimensional transition
state theory (TST) has been presented in Sec.III. We work out the
rate expression and the Wolynes factor in the appropriate limits.
The paper is concluded in Sec.IV.

\section{{quantum Langevin equation}}

We consider a particle of unit mass coupled to a medium comprised
of a set of harmonic oscillators with frequency $\omega_i$. This
is described by the following Hamiltonian:

\begin{equation}\label{2.1}
\hat{H} = \frac{\hat{p}^2}{2} + V(\hat{q}) + \sum^N_{i=1} \left \{
\frac{\hat{p}_i^2}{2} + \frac{1}{2} \left(\omega_i \hat{x}_i -
\frac{c_i}{\omega_i} \hat{q} \right)^2 \right \}
\end{equation}

Here $\hat{q}$ and $\hat{p}$ are co-ordinate and momentum
operators of the particle and the set $\{ \hat{x}_i,\hat{p}_i \}$
is the set of co-ordinate and momentum operators for the
reservoir oscillators coupled linearly to the system through
their coupling coefficients $c_i$. The potential $V(\hat{q})$ is
due to the external force field for the Brownian particle. The
co-ordinate and momentum operators follow the usual commutation
relations [$\hat{q}, \hat{p}$] = $i \hbar$ and [$\hat{x}_i,
\hat{p}_j$] = $i \hbar \delta_{ij}$. Here masses have been
assumed to be unity.

Eliminating the reservoir degrees of freedom in the usual way we
obtain the operator Langevin equation for the particle,

\begin{equation}\label{2.2}
\ddot{ \hat{q} } (t) + \int_0^t dt'  \gamma (t-t')\; \dot{ \hat{q}
} (t') + V' ( \hat{q} ) = \hat{F} (t) \; \; ,
\end{equation}

where the noise operator $\hat{F} (t)$ and the memory kernel
$\gamma (t)$ are given by

\begin{equation}\label{2.3}
\hat{F} (t) = \sum_j \left [ \left \{ \frac{\omega_j^2}{c_j}
\hat{x}_j (0) - \hat{q} (0) \right \} \frac{c_j^2}{\omega_j^2}
\cos \omega_j t + \frac{c_j}{\omega_j} \hat{p}_j (0) \sin
\omega_j t \right ]
\end{equation}

and

\begin{equation}\label{2.4}
\gamma (t) = \sum_{j=1}^N \frac{c_j^2}{\omega_j^2} \cos \omega_j t
\end{equation}

The Eq.(\ref{2.2}) is the well known exact quantized operator
Langevin equation for which the noise properties of $\hat{F} (t)$
can be derived by using a suitable initial canonical distribution
of the bath co-ordinate and momentum operators at $t=0$ as
follows;

\begin{eqnarray}
\langle \hat{F} (t) \rangle_{QS} &=& 0 \label{2.5} \\
\frac{1}{2} \{ \langle\hat{F}(t)\hat{F}(t^\prime)\rangle_{QS}+
\langle\hat{F}(t^\prime)\hat{F}(t)\rangle_{QS} \} &=& \frac{1}{2}
\sum_{j=1}^N \hbar \omega_j \left(\coth \frac{\hbar\omega_j}{2
k_B T}\right)\frac{c_j^2}{\omega_j^2}\; \cos
\omega_j(t-t^\prime)\label{2.6}
\end{eqnarray}

where $\langle...\rangle_{QS}$ refers to quantum statistical
average on bath degrees of freedom and is defined as

\begin{equation}\label{2.7}
\langle \hat{O} \rangle_{QS} = \frac{{\rm Tr} \hat{O} \exp
(-\hat{H}_{{\rm bath}}/k_BT)}{{\rm Tr}\exp (-\hat{H}_{{\rm
bath}}/k_BT)}
\end{equation}

for any operator $\hat{O}(\{(\omega_j^2/c_j)\hat{x}_j -
\hat{q}\},\{\hat{p}_j\})$ where $\hat{H}_{{\rm bath}}
(\sum^N_{i=1} [\hat{p}_i^2/2 + 1/2 (\omega_i \hat{x}_i -
\frac{c_i}{\omega_i} \hat{q} )^2])$ at $t=0$. By Trace we mean the
usual quantum statistical average. Eq.(\ref{2.6}) is the
fluctuation-dissipation relation with the noise operators ordered
appropriately in the quantum mechanical sense.

To construct a c-number Langevin equation \cite{db1,bk1} we
proceed from Eq.(\ref{2.2}). We carry out a {\it quantum
mechanical average} of Eq.(\ref{2.2})

\begin{equation}\label{2.8}
\langle \ddot{ \hat{q} } (t) \rangle + \int_0^t dt' \gamma (t-t')
\langle \dot{ \hat{q} } (t') \rangle + \langle V' ( \hat{q} )
\rangle = \langle \hat{F} (t) \rangle
\end{equation}

where the quantum mechanical average $\langle \ldots \rangle$ is
taken over the initial product separable quantum states of the
particle and the bath oscillators at $t=0$, $| \phi \rangle \{ |
\alpha_1 \rangle | \alpha_2 \rangle \ldots | \alpha_N \rangle \}
$. Here $| \phi \rangle$ denotes any arbitrary initial state of
the particle and $| \alpha_i \rangle$ corresponds to the initial
coherent state of the $i$-th bath oscillator. $|\alpha_i \rangle$
is given by $|\alpha_i \rangle = \exp(-|\alpha_i|^2/2)
\sum_{n_i=0}^\infty (\alpha_i^{n_i} /\sqrt{n_i !} ) | n_i \rangle
$, $\alpha_i$ being expressed in terms of the mean values of the
shifted co-ordinate and momentum of the $i$-th oscillator,
$\{(\omega_i^2/c_i)\langle \hat{x}_i (0) \rangle - \langle
\hat{q}(0) \rangle\} = ( \sqrt{\hbar} /2\omega_i) (\alpha_i +
\alpha_i^\star )$ and $\langle \hat{p}_i (0) \rangle = i
\sqrt{\hbar\omega_i/2 } (\alpha_i^\star - \alpha_i )$,
respectively. It is important to note that $\langle \hat{F} (t)
\rangle$ of Eq.(\ref{2.8}) is a classical-like noise term which,
in general, is a non-zero number because of the quantum mechanical
averaging and is given by $(\langle \hat{F}(t) \rangle \equiv
f(t))$;

\begin{equation}\label{2.9}
f(t) = \sum_j \left [ \left \{ \frac{\omega_j^2}{c_j} \langle
\hat{x}_j (0)\rangle - \langle\hat{q} (0)\rangle \right \}
\frac{c_j^2}{\omega_j^2} \cos \omega_j t + \frac{c_j}{\omega_j}
\langle\hat{p}_j (0)\rangle \sin \omega_j t \right ]
\end{equation}

It is convenient to rewrite the $c$-number equation (\ref{2.8}) as
follows;

\begin{equation}\label{2.10}
\langle \ddot{ \hat{q} } (t) \rangle + \int_0^t dt' \gamma (t-t')
\langle \dot{ \hat{q} } (t') \rangle + \langle V' ( \hat{q} )
\rangle = f (t)
\end{equation}

To realize $f(t)$ as an effective c-number noise we now introduce
the ansatz that the momenta $\langle \hat{p}_j (0) \rangle$ and
the shifted co-ordinates
$\{(\omega_j^2/c_j)\langle\hat{x}_j(0)\rangle -
\langle\hat{q}(0)\rangle\}$, $\{\hat{p}_j\}$ of the bath
oscillators are distributed according to a canonical distribution
of Gaussian form as

\begin{equation}\label{2.11}
{P}_j = {N} \exp \left\{ -\;\frac{\langle \hat{p}_j(0) \rangle^2
+ \frac{c_j^2}{\omega_j^2} \{ \frac{\omega_j^2}{c_j}\langle
\hat{x}_j (0) \rangle - \langle \hat{q} (0) \rangle \}^2 }{ 2
\hbar \omega_j \left( \bar{n}_j(\omega_j) + \frac{1}{2} \right) }
\right\}
\end{equation}

so that for any function of the quantum mechanical mean values
$O_j\{\langle\hat{p}_j(0)\rangle,\\(({\omega_j^2}/{c_j})\langle\hat{x}_j
(0) \rangle  - \langle \hat{q} (0) \rangle )\}$ of the bath the
statistical average $\langle \ldots \rangle_S$ is

\begin{eqnarray}\label{2.12}
\langle O_j \rangle_S & = &\int O_j \;{P}_j \; d\langle \hat{p}_j
(0) \rangle \;d \{ (\omega_j^2/c_j)\langle \hat{x}_j (0) \rangle -
\langle \hat{q} (0) \rangle \}
\end{eqnarray}

\noindent Here $\bar{n}_j$ indicates the average thermal photon
number of the $j$-th oscillator at temperature $T$ and
$\bar{n}_j(\omega_j) = 1/[\exp \left ( \hbar \omega_j/k_BT \right
) - 1]$ and ${N}$ is the normalization constant.

The distribution (\ref{2.11}) and the definition of statistical
average (\ref{2.12}) imply that $f(t)$ must satisfy

\begin{equation}\label{2.13}
\langle f(t) \rangle_S = 0
\end{equation}

and

\begin{equation}\label{2.14}
\langle f(t)f(t^{\prime})\rangle_S = \frac{1}{2} \sum_j \hbar
\omega_j \left ( \coth \frac { \hbar \omega_j }{ 2k_BT } \right )
\frac{c_j^2}{\omega_j^2}\;\cos \omega_j (t - t^{\prime})
\end{equation}

\noindent That is, $c$-number noise $f(t)$ is such that it is
zero-centered and satisfies the standard fluctuation-dissipation
relation (FDR) as expressed in Eq.(\ref{2.6}). It is important to
emphasize that the ansatz (\ref{2.11}) is a canonical Wigner
distribution for a shifted harmonic oscillator \cite{wig,hil} and
an exact solution of Wigner equation, which remains always a
positive definite function. A special advantage of using this
distribution is that it remains valid as pure state non-singular
distribution function at $T = 0$. Furthermore, this procedure
allows us to {\it bypass the operator ordering} prescription of
Eq.(\ref{2.6}) for deriving the noise properties of the bath in
terms of fluctuation-dissipation relation and to identify $f(t)$
as a classical looking noise with quantum mechanical content.

We now return to Eq.(\ref{2.10}) to add the force term
$V^\prime(\langle\hat{q} \rangle)$ on both sides of
Eq.(\ref{2.10}) and rearrange it to obtain \cite{db1,bk1}

\begin{eqnarray}
\dot q &=& p
\label{2.15}\\
\dot p &=& - \int_0^t dt^\prime \gamma (t - t^\prime) p(t^\prime)
- V^\prime (q) + f(t) + Q \label{2.16}
\end{eqnarray}

where we put $\langle\hat{q}(t)\rangle = q(t)$ and
$\langle\hat{p}(t)=p(t)$ for notational convenience and

\begin{equation}\label{2.17}
Q = V^\prime(\langle\hat{q}\rangle) - \langle V^\prime(\hat{q})
\rangle
\end{equation}

represents the quantum correction due to the system degrees of
freedom. Eq.(\ref{2.16}) offers a simple interpretation. This
implies that the quantum Langevin equation is governed by a
$c$-number quantum noise $f(t)$ originating from the heat bath
characterized by the properties (\ref{2.13}) and (\ref{2.14}) and
a quantum fluctuation term $Q$ \cite{db1,bk1,sm,akp}
characteristic of the non-linearity of the potential \cite{tani}.

Referring to the quantum nature of the system in the Heisenberg
picture, we write \cite{sm,akp}.

\begin{eqnarray}
\hat{q} (t) &=& q + \delta\hat{q} \label{2.18}  \\
 \hat{p} (t) &=& p + \delta\hat{p}\label{2.19}
\end{eqnarray}

where $\langle\hat{q}\rangle(=q)$ and $\langle\hat{p}\rangle(=p)$
are the quantum-mechanical averages and $\delta\hat{q}$,
$\delta\hat{p}$ are the operators. By construction
$\langle\delta\hat{q}\rangle$ and $\langle\delta\hat{p}\rangle$
are zero and $[\delta\hat{q},\delta\hat{p}] = i\hbar$. Using
Eqs.(\ref{2.18}) and (\ref{2.19}) in $\langle V^{\prime} (\hat{q})
\rangle$ and a Taylor series expansion around
$\langle\hat{q}\rangle$ it is possible to express $Q$ as

\begin{equation}\label{2.20}
Q(q, \langle\delta\hat{q}^n\rangle) = -\sum_{n \ge 2}
\frac{1}{n!} V^{(n+1)} (q) \langle\delta\hat{q}^n\rangle
\end{equation}

Here $V^{(n)}(q)$ is the n-th derivative of the potential $V(q)$.
For example, the second order $Q$ is given by $Q = -\frac{1}{2}
V^{\prime\prime\prime} (q) \langle \delta \hat{q}^2 \rangle$. The
calculation of $Q$ \cite{db1,bk1,sm,akp} therefore rests on
quantum correction terms, $\langle \delta \hat{q}^n \rangle$
which are determined by solving the quantum correction equations
as discussed in the following subsection.

\subsection{Quantum correction equations}

We now return to operator equation (\ref{2.2}) and put
(\ref{2.18}) and (\ref{2.19}) and use of Eq.(\ref{2.10}) to
obtain the following operator equations

\begin{equation}\label{2.21}
\dot{\delta\hat{q}}=\delta\hat{p}
\end{equation}

\newpage

\begin{eqnarray}
\dot{\delta\hat p} + \int_0^t \gamma(t-t^\prime) \;\delta \hat
p(t^\prime)\; dt^\prime &+& V^{\prime \prime} (q) \;\delta \hat q
+ \sum_{n\geq 2}\frac{1}{n!} \;V^{(n+1)}(q)\;\left( \delta \hat
q^n-\langle\delta \hat q^n\rangle\right)\nonumber\\
&=&\hat{F}(t)-f(t)\label{2.22}
\end{eqnarray}

Eqs.(2.21-2.22) form the key element for calculation of quantum
mechanical correction (\ref{2.20}) due to nonlinearity of the
system potential. Depending on nonlinearity of the potential and
memory kernel we consider the following cases separately.

\subsubsection{\bf {\textit{Arbitrary memory kernel and harmonic
potential}}}

We consider the friction kernel to be arbitrary but decaying and
the potential as harmonic for which the derivatives of $V(q)$
higher than second vanishes. After quantum mechanical averaging
over the initial bath states $\prod_{i=1}^{\infty} \left \{
|\alpha_i (0) \rangle \right\}$ only, the
Eqs.(\ref{2.21}-\ref{2.22}) yield

\begin{equation}\label{2.23}
\dot{\delta\hat{q}}=\delta\hat{p}
\end{equation}

\begin{equation}\label{2.24}
\dot{\delta \hat p}= - \int_0^t \gamma(t-t^\prime) \;\delta \hat
p(t^\prime)\; dt^\prime -\omega^2 \delta \hat{q}
\end{equation}

where $\omega^2=V^{\prime\prime}(q)$, a constant and $\omega$ is
the frequency of the harmonic oscillator. The above operator
equation can be solved exactly by Laplace transform technique to
obtain

\begin{equation}\label{2.25}
\delta {\hat q}(t)= \delta {\hat q}(0)\left( 1 -\omega^2 \int_0^t
\;C_p(t^\prime) dt^\prime \right)+\delta \hat p(0)\; C_p(t)
\end{equation}

where $C_p(t)$ is obtained as the inverse Laplace transform of
$\widetilde{C_p}(s)$ given by

\begin{equation}\label{2.26}
\widetilde{C_p}(s)=\frac{1}{s^2+s\; \widetilde{\gamma}(s) +
\omega^2}
\end{equation}

with $\widetilde{\gamma}(s)$ as

\begin{equation}\label{2.27}
\widetilde{\gamma}(s)=\int_0^\infty \gamma(t) e^{-s t} dt
\end{equation}

is the Laplace transform of the friction kernel $\gamma(t)$. After
squaring and quantum mechanical averaging over arbitrary initial
system state $|\phi\rangle$, Eq.(\ref{2.25}) yields

\begin{eqnarray}
\langle \delta \hat q^2(t) \rangle &=& \langle \delta \hat
q^2(0)\; \rangle \;C_q^2(t)+\langle \delta \hat
p^2(0)\; \rangle \;C_p^2(t)\nonumber\\
&+& C_p(t)\; C_q(t)\; \{\langle \delta \hat p(0)\; \delta \hat
q(0)\rangle+\langle\delta \hat q(0)\; \delta \hat p(0)
\rangle\}\label{2.28}
\end{eqnarray}

with $C_q(t)=1-\omega^2\int_0^t C_p(t^\prime)dt^\prime$.

\subsubsection{\bf {\textit{Exponential memory kernel and arbitrary
potential}}}

This corresponds to a commonly occurring situation for which the
bath modes are assumed to follow a Lorentzian distribution
characterized by a density function $\rho(\omega)$ such that for
Eq.(\ref{2.4}) in the continuum limit we write

\begin{equation}\label{2.29}
\frac{c^2(\omega)}{\omega^2}\;\rho(\omega)=\frac{2}{\pi}\;
\frac{\Gamma}{1+\omega^2\tau_c^2}
\end{equation}

where $\Gamma$ is the dissipation constant in the Markovian limit
and $\tau_c$ refers to correlation time. The memory kernel is
exponential and $\gamma(t)$ is given by

\begin{equation}\label{2.30}
\gamma(t)=\frac{\Gamma}{\tau_c}\;e^{-|t|/\tau_c}
\end{equation}

In this case the quantum correction equations (\ref{2.21}) and
(\ref{2.22}) after quantum mechanical averaging over bath states
may be rewritten

\begin{equation}
\dot{\delta\hat{q}}=\delta\hat{p}\label{2.31}
\end{equation}

\begin{eqnarray}\label{2.32}
\dot{\delta\hat p} = - V^{\prime \prime} (q) \;\delta \hat q -
\sum_{n\geq 2}\frac{1}{n!} \;V^{(n+1)}(q)\;\left( \delta \hat
q^n-\langle\delta \hat q^n\rangle\right)+\delta\hat{z}
\end{eqnarray}

\begin{equation}\label{2.33}
\dot{\delta \hat{z}} = -\;\frac{\Gamma}{\tau_c}\;\delta \hat{p}\;
- \frac{1}{\tau_c}\; \delta \hat{z}
\end{equation}

where we have introduced an auxiliary operator $\delta\hat{z}$ to
bypass the convolution integral in Eq.(\ref{2.22}). Making use of
the Eqs.(\ref{2.31}-\ref{2.33}) we derive the quantum correction
equations upto second order to obtain

\begin{equation}\label{2.34}
\langle \dot{\delta \hat{q}}^2 \rangle = \langle \delta\hat{q}
\delta\hat{p} + \delta\hat{p} \delta\hat{q} \rangle
\end{equation}

\begin{equation}\label{2.35}
\langle \dot{\delta\hat{q} \delta\hat{p} + \delta\hat{p}
\delta\hat{q}}\;\; \rangle = 2\;\langle \delta\hat{p}^2 \rangle -
2\;V^{\prime\prime}(q) \langle \delta\hat{q}^2 \rangle + \langle
\delta\hat{q} \delta\hat{z} + \delta\hat{z} \delta\hat{q} \rangle
- V^{\prime\prime\prime}(q)\langle \delta\hat{q}^3 \rangle
\end{equation}

\begin{equation}\label{2.36}
\langle \dot{\delta\hat{p}}^2 \rangle =
-V^{\prime\prime}(q)\langle \delta\hat{q} \delta\hat{p} +
\delta\hat{p} \delta\hat{q} \rangle + \langle \delta\hat{p}
\delta\hat{z} + \delta\hat{z} \delta\hat{p} \rangle -
\frac{1}{2}V^{\prime\prime\prime}(q) \langle \delta\hat{p}
\delta\hat{q}^2 + \delta\hat{q}^2 \delta\hat{p} \rangle
\end{equation}

\begin{eqnarray}
\langle \dot{\delta\hat{p} \delta\hat{z} + \delta\hat{z}
\delta\hat{p}}\;\; \rangle = &-& \; V^{\prime\prime}(q) \langle
\delta\hat{q} \delta\hat{z} + \delta\hat{z} \delta\hat{q} \rangle-
\frac{1}{2}V^{\prime\prime\prime}(q) \langle \delta\hat{q}^2
\delta\hat{z} + \delta\hat{z} \delta\hat{q}^2 \rangle \nonumber\\
&-&\frac{2\Gamma}{\tau_c} \;\langle \delta\hat{p}^2 \rangle -
\frac{1}{\tau_c}\; \langle \delta\hat{p} \delta\hat{z} +
\delta\hat{z} \delta\hat{p} \rangle + 2\;\langle \delta\hat{z}^2
\rangle \label{2.37}
\end{eqnarray}

\begin{equation}\label{2.38}
\langle \dot{\delta\hat{q} \delta\hat{z} + \delta\hat{z}
\delta\hat{q}}\;\; \rangle = \langle \delta\hat{p} \delta\hat{z}
+ \delta\hat{z} \delta\hat{p} \rangle - \frac{\Gamma}{\tau_c}\;
\langle \delta\hat{q} \delta\hat{p} + \delta\hat{p} \delta\hat{q}
\rangle - \frac{1}{\tau_c}\; \langle \delta\hat{q} \delta\hat{z}
+ \delta\hat{z} \delta\hat{q} \rangle
\end{equation}

\begin{equation}\label{2.39}
\langle \dot{\delta\hat{z}}^2 \rangle = -\frac{\Gamma}{\tau_c}\;
\langle \delta\hat{p} \delta\hat{z} + \delta\hat{z} \delta\hat{p}
\rangle - \frac{2}{\tau_c}\;\langle \delta\hat{z}^2 \rangle
\end{equation}

Discarding the third and higher order terms in
Eqs.(\ref{2.34}-\ref{2.39}) we obtain a set of closed equations
which can be solved numerically along with Eqs.(\ref{2.15}) and
(\ref{2.16}), using suitable initial conditions for calculating
the leading order (second order) contribution to quantum
corrections in $Q$.A standard choice of initial conditions
\cite{sm,akp} corresponding to minimum uncertainty states is
$\langle \delta \hat{q}^2(0) \rangle = \hbar/2 \omega$, $\langle
\delta \hat{p}^2(0) \rangle = \hbar \omega/2$ and $\langle \delta
\hat{q}(0) \delta \hat{p}(0) + \delta \hat{p}(0) \delta
\hat{q}(0) \rangle = \hbar$ with the other moments
(Eqs.(\ref{2.37}-\ref{2.39})) being set at zero. The procedure
may be extended to include higher order quantum effects without
difficulty and may be easily adopted for numerical simulation of
quantum Brownian motion \cite{bk1}.

\subsection{Calculation of quantum statistical averages}

Summarizing the discussions of the last two sections $A$ and $B$
we now see that the Langevin dynamics in c-numbers can be
calculated for a stochastic process by solving Eqs.(\ref{2.15})
and (\ref{2.16}) for quantum mechanical mean values simultaneously
with quantum correction equations which describe quantum
fluctuation around these mean values. In principle for nonlinear
system the equations for quantum corrections constitute an
infinite set of hierarchy which must be truncated after a desired
order, in practice, to make the system of equations closed. Care
must be taken to distinguish among the three averages, the quantum
mechanical mean $\langle\hat{O}\rangle(=O)$, statistical average
over quantum mechanical mean $\langle O \rangle_S$ and the usual
quantum statistical average $\langle\hat{O}\rangle_{QS}$ as
discussed in Sec.II. To illustrate the relation among them let us
calculate, for example, the quantum statistical averages
$\langle\hat{q}\rangle_{QS}$, $\langle\hat{q}^2\rangle_{QS}$ and
$\langle\hat{q}^2\hat{p}\rangle_{QS}$. By (\ref{2.18}) and
(\ref{2.19}) we write

\begin{eqnarray}
\hat{q}&=&q+\delta\hat{q}\nonumber\\
\langle\hat{q}\rangle_{QS}&=&\langle {q + \delta\hat{q}}
\rangle_{QS}\label{2.40}\\
&=& \langle q \rangle_{S}+\langle \langle \delta \hat{q} \rangle
\rangle_S=\langle q \rangle_{S}\label{2.41}
\end{eqnarray}

Again

\begin{eqnarray}
\langle\hat{q}^2\rangle_{QS}&=&\langle(q+\delta\hat{q})^2\rangle_{QS}\nonumber\\
&=&\langle q^2 \rangle_S+\langle \langle \delta \hat{q}^2 \rangle
\rangle_S\label{2.42}
\end{eqnarray}

In the case of harmonic potential, $\langle\delta
\hat{q}^2\rangle$ as given by (\ref{2.28}) is independent of $q$
or $p$ so that one may simplify (\ref{2.42}) further as

\begin{equation}\label{2.43}
\langle\hat{q}^2\rangle_{QS}=\langle q^2 \rangle_S+\langle \delta
\hat{q}^2 \rangle
\end{equation}

In Ref. \cite{db1} the explicit exact expressions for $\langle
\hat{q}^2\rangle_{QS}$ and $\langle\hat{p}^2\rangle_{QS}$ have
been derived for harmonic oscillator and they are found to be in
exact agreement with those of Grabert \textit{et al} \cite{bat}.
For anharmonic potential, however, one must have to use
(\ref{2.42}) to carry out further the statistical average over
$\langle\delta\hat{q}^2\rangle$ \textit{i.e}
$\langle\langle\delta\hat{q}^2\rangle\rangle_S$, since it is a
function of stochastic variables $q$ and $p$ according to quantum
correction equations. Furthermore we consider
$\langle\hat{q}^2\hat{p}\rangle_{QS}$

\begin{eqnarray}
\langle\hat{q}^2\hat{p}\rangle_{QS}&=&\langle
(q+\delta\hat{q})^2(p+\delta\hat{p}) \rangle_{QS}\nonumber\\
&=&\langle q^2p\rangle_S+\langle p \;\langle \delta\hat{q}^2
\rangle\rangle_S+\langle \langle \delta\hat{q}^2 \;\delta\hat{p}
\rangle\rangle_S+2 \;\langle\; q \;\langle
\delta\hat{q}\;\delta\hat{p}\rangle\rangle_S\label{2.44}
\end{eqnarray}

The essential element of the present approach is thus expressing
the quantum statistical average as the sum of statistical
averages of set of functions of quantum mechanical mean values
and dispersions. Langevin dynamics being coupled to quantum
correction equations, the quantum mechanical mean values as well
as the dispersions are computed simultaneously for each
realization of the stochastic path \cite{db1}. A statistical
average implies the averaging over many such paths similar to
what is done to calculate statistical averaging by solving
classical Langevin equation. Before leaving this section we
mention a few pertinent points.

First, the distinction between the ensemble averaging by the
present procedure and by the standard approach using Wigner
function is now clear. From Eq.(\ref{2.44}) we note that, for
example,

\begin{equation}\label{2.45}
\langle \hat{q}^2 \hat{p} \rangle_{QS}=\int q^2 p\; W(q,p)\neq
\langle q^2 p \rangle_S
\end{equation}

where $W(q,p)$ is the Wigner function for system operators. (This
is not to be confused with the Wigner function we introduced in
(\ref{2.11}) for the bath oscillators.)

Second, our formulation of the Langevin equation coupled to
quantum correction equations belongs to quantum stochastic process
derived by c-number  noise, which is classical-like in form. Its
numerical solutions can be obtained in the same way as one
proceeds in a classical theory \cite{db1}.

Third, quantum nature of the dynamics appears in two different
ways. The heat bath is quantum mechanical in character whose noise
properties are expressed through quantum fluctuation-dissipation
relation. The nonlinearity of the system potential, on the other
hand, gives rise to quantum correction terms. Thus the classical
Langevin equation can be easily recovered (i) in the limit
$\hbar\omega\ll k_BT$ to be applied in the Eq.(\ref{2.14}) so
that one obtains the classical fluctuation-dissipation relation
and (ii) if the quantum dispersion term $Q$ vanishes.

\subsection{c-number Hamiltonian and normal mode description}

The c-number Hamiltonian corresponding to Langevin equation
(\ref{2.15}) and (\ref{2.16}) is given by

\begin{eqnarray}
H = \frac{p^2}{2} + \left[ V(q) + \sum_{n \geq 2} \frac{1}{n!}
V^{(n)}(q) \langle {\delta \hat q^n} \rangle \right]
+\sum^N_{i=1} \left \{ \frac{p_i^2}{2} + \frac{1}{2} \left(
\omega_i x_i - \frac{c_i}{\omega_i} q \right)^2 \right
\}\label{2.46}
\end{eqnarray}

Note that the above Hamiltonian is different from our starting
Hamiltonian operator (\ref{2.1}) because of the c-number nature of
(\ref{2.46}). $\{x_i,p_i\}$ are the quantum mean value of the
co-ordinate and the momentum operators of the bath oscillators.

The spectral density function is defined as

\begin{equation}\label{2.47}
J(\omega)=\frac{\pi}{2} \sum_{i=1}^N \frac{c_i^2}{\omega_i}\;
\delta(\omega-\omega_i)
\end{equation}

We now assume that at $q=0$, the potential $V(q)$ has a barrier
with height $V^\ddag$ such that a harmonic approximation around
$q=0$ leads to

\begin{equation}\label{2.48}
V(q) = V^\ddag - \frac{1}{2} \omega_b^2 q^2 + V_2(q)
\end{equation}

where $\omega_b^2 = V''(q)\mid_{q=0}$, refers to the second
derivative of the potential $V(q)$. $\omega_b$ is the frequency at
the barrier top and $V_2(q)$ is the non-linear part of the
classical potential and is given by
$V_2=\sum_{n\ge3}\frac{1}{n!}\frac{\partial^nV(q)}{\partial
q^n}\mid_{q=0}q^n$. With Eq.(\ref{2.48}) the quantum correction
part in the Hamiltonian Eq.(\ref{2.46}) becomes

\begin{equation}
\sum_{n\geq 2} \frac{1}{n!} V^{(n)}(q) \langle \overline{\delta
\hat q^n} \rangle = -\frac{\omega_b^2}{2} B_2 + V_3(q)\label{2.49}
\end{equation}

where $B_n=\langle \overline{\delta \hat q^n} \rangle ;
V_3=\sum_{n \geq 2} \frac{B_n}{n!}
\frac{\partial^nV_2(q)}{\partial q^n}$. Note that we have
introduced an approximation by putting a bar over quantum
dispersion $\langle \delta \hat{q}^n \rangle$ to indicate its
time average since it is sufficient to consider the energy loss of
the system mode averaged over one round trip time, \textit{i.e.},
the time required to traverse from one turning point of the
potential well to another and back. Putting (\ref{2.48}) and
(\ref{2.49}) in the Hamiltonian (\ref{2.46}) we obtain

\begin{equation}\label{2.50}
H = H_0 + V_N (q)
\end{equation}

where we have decomposed the Hamiltonian in the harmonic part
$H_0$ and the anharmonic part $V_N(q)$ as

\begin{equation}\label{2.51}
H_0 =\left[ \frac{p^2}{2} + \sum_i \frac{p_i^2}{2} \right] +
\left[ V_1^\ddag - \frac{1}{2} \omega_b^2 q^2 + \sum_i \frac{1}{2}
\left( \omega_i x_i - \frac{c_i}{\omega_i} q \right)^2 \right]
\end{equation}

and

\begin{eqnarray}
V_N(q) & = & V_2(q)+V_3(q)\label{2.52}\\
and\;\;\;\;\;
 V_1^\ddag & = & V^\ddag - \frac{B_2}{2} \omega_b^2\nonumber
\end{eqnarray}

$V_2(q)$ and $V_3(q)$ are therefore classical and quantum
anharmonic contributions to total anharmonic part of the
Hamiltonian. The separability of the c-number Hamiltonian in the
quadratic and nonlinear parts allows us to make a normal mode
transformation to convert the quadratic Hamiltonian into a
diagonal form. The method of normal mode analysis has been used
extensively by Pollak and co-workers \cite{p1}.

Following Pollak, we diagonalize the force constant matrix $T$ of
the Hamiltonian (\ref{2.51}) with the matrix $U$

\begin{equation}\label{2.53}
U T = \lambda^2 U
\end{equation}

where $U$ provides the transformation from old co-ordinates to the
normal co-ordinates

\begin{eqnarray}
\left(
\begin{array}{cc}
{\rho}\\
{y_1}\\
{y_2}\\
{.}\\
{.}\\
{y_N}
\end{array}
\right) = U \left(
\begin{array}{cc}
{q}\\
{x_1}\\
{x_2}\\
{.}\\
{.}\\
{x_N}
\end{array}
\right)\label{2.54}
\end{eqnarray}

The c-number Hamiltonian of the unstable normal co-ordinate is
given by

\begin{equation}\label{2.55}
H_0 = \frac{1}{2} \dot \rho^2 + V_1^\ddag - \frac{1}{2}
\lambda_b^2 \rho^2 + \sum_{i=1}^N \frac{1}{2} \left( \dot y_i^2 +
\lambda_i^2 y_i^2 \right)
\end{equation}

The eigenvalues $\lambda_i^2$ and $\lambda_b^2$ are expressible
in terms of the coupling constant of the system and the bath
implicitly as follows:

\begin{eqnarray}
\lambda_b^2 & = & \omega_b^2\left[1 + \sum_{j=1}^N
\frac{c_j^2}{\omega_j^2 (\omega_j^2 + \lambda_b^2)} \right]^{-1} \label{2.56}\\
\lambda_i^2 & = & - \;\omega_b^2 \left[1 + \sum_{j=1}^N
\frac{c_j^2}{\omega_j^2 (\omega_j^2 - \lambda_i^2)}
\right]^{-1},\;\;\; i=1,2...N \label{2.57}
\end{eqnarray}

where (\ref{2.56}) and (\ref{2.57}) correspond to normal mode
frequencies of the unstable mode and the i-th bath oscillator
respectively.

The transformation (\ref{2.54}) implies

\begin{equation}\label{2.58}
q = u_{00}\; \rho +\sum_{j=1}^N u_{j0} \;y_{j}
\end{equation}

and it has been shown \cite{p1} that $u_{00}$ and $u_{j0}$ may be
expressed as

\begin{equation}\label{2.59}
u_{00}^2 = \left[ 1 + \sum_{j=1}^N \frac{c_j^2}{ (\omega_j^2 +
\lambda_b^2)^2} \right]^{-1}
\end{equation}

and

\begin{equation}\label{2.60}
u_{j0}^2 = \left[ 1 + \sum_{j=1}^N \frac{c_j^2}{ (\lambda_j^2 -
\omega_j^2)^2} \right]^{-1}
\end{equation}

Making use of the spectral density function (\ref{2.47}) and
Laplace transformation of $\gamma(t)$, Eq.(\ref{2.56}) and
Eq.(\ref{2.59}) may be written in the continuum limit as

\begin{equation}\label{2.61}
\lambda_b^2 = \frac{\omega_b^2}{1 + \widetilde{ \gamma}
(\lambda_b)/\lambda_b}
\end{equation}

and

\begin{equation}\label{2.62}
u_{00}^2 = \left[ 1 + \frac{2}{\pi} \int_0^\infty d\omega
\frac{\omega\;J(\omega) \; }{(\lambda_b^2 + \omega^2)^2}
\right]^{-1}
\end{equation}

The two important identities in relation to orthogonal
transformation matrices and the associated frequencies may be
noted here for the dynamics at the barrier top and at the bottom
of the well;

\begin{equation}\label{2.63}
\omega_b^2 \prod_{i=1}^N \omega_i^2 = \lambda_b^2 \prod_{i=1}^N
\lambda_i^2
\end{equation}

and

\begin{equation}\label{2.64}
\omega_0^2 \prod_{i=1}^N \omega_i^2 = \lambda_0^2 \prod_{i=1}^N
\Lambda_i^2
\end{equation}

Here $\omega_0$ and $\lambda_0$ are the frequencies of the system
at the bottom of the well in the original co-ordinate and normal
co-ordinate respectively. Similarly $\Lambda_i$ corresponds to the
normal mode frequencies of the bath oscillators corresponding to a
normal mode Hamiltonian at the bottom of the well,

\begin{equation}
H_0^\prime = \frac{1}{2} \dot \rho^{\prime 2} + \frac{1}{2}
\lambda_0^2 \rho^{\prime 2} + \left\{ \sum_{i=1}^N \frac{1}{2}
\dot y_i^{\prime 2} + \frac{1}{2} \Lambda_i^2\; y_i^{\prime 2}
\right\}\label{2.65}
\end{equation}

Here $\rho^\prime$ and $\dot \rho^{\prime}$ are coordinate and
momentum of system mode respectively and $y_{i}^{\prime}$ and
$\dot y_{i}^{\prime}$ are coordinate and momentum of $i^{th}$
bath oscillator respectively at the bottom of the well in the
normal coordinates. $\lambda_0$ and $\Lambda_i$ are given by
\cite{p1}

\begin{eqnarray}
\lambda_0^2 & = & \omega_0^2\left[1 + \sum_{j=1}^N
\frac{c_j^2}{\omega_j^2 (\omega_j^2 - \lambda_0^2)} \right]^{-1} \label{2.66}\\
\Lambda_i^2 & = & \omega_0^2\left[1 + \sum_{j=1}^N
\frac{c_j^2}{\omega_j^2 (\omega_j^2 - \Lambda_j^2)}
\right]^{-1},\;\;\; i=1,2...N \label{2.67}
\end{eqnarray}

\section{c-number quantum version of
multidimensional TST}

To begin with we consider the particle to be trapped in a well
described by a potential $V(q)$. In the normal mode description
of $(N+1)$ oscillators according to the Hamiltonian (\ref{2.55})
the bath modes and the system mode are uncoupled. Considering the
unstable reaction co-ordinate to be thermalized according to the
Wigner thermal canonical distribution \cite{wig,hil} of $N$
uncoupled harmonic oscillators plus one inverted we have

\begin{equation}\label{3.1}
P_{eq} = z^{-1} \exp \left[ - \;\frac{\frac{1}{2} \dot \rho^2 +
V_1^\ddag - \frac{1}{2} \lambda_b^2 \rho^2}{\hbar \lambda_0
(\overline{n}_0(\lambda_0) + \frac{1}{2})} \right] \prod_{i=1}^N
\exp \left[ - \;\frac{\frac{1}{2} \dot y_i^2 + \frac{1}{2}
\lambda_i^2 y_i^2}{\hbar \lambda_i (\overline{n}_i(\lambda_i) +
\frac{1}{2})} \right]
\end{equation}

$z$ is the normalization constant. As usual this can be
calculated using the distribution function inside the reactant
well. For this it is necessary to consider the normal mode
Hamiltonian at the bottom of the well expressed as $H_0^{\prime}$
in the Eq.(\ref{2.65}). The corresponding distribution in the well
is

\begin{equation}\label{3.2}
P_{eq} = z^{-1} \exp \left[ -\; \frac{\frac{1}{2} \dot
\rho^{\prime 2} + \frac{1}{2} \lambda_0^2 \rho^{\prime 2}}{\hbar
\lambda_0 (\overline{n}_0(\lambda_0) + \frac{1}{2})} \right]
\prod_{i=1}^N \exp \left[  -\; \frac{\frac{1}{2} \dot y_i^{\prime
2} + \frac{1}{2} \Lambda_i^2 y_i^{\prime 2}}{\hbar \Lambda_i
(\overline{n}_i(\Lambda_i) + \frac{1}{2})} \right]
\end{equation}

which can be normalized to obtain

\begin{equation}\label{3.3}
z^{-1} = \frac{\lambda_0}{ 2 \pi \hbar \lambda_0
(\overline{n}_0(\lambda_0) + \frac{1}{2})} \; \prod_{i=1}^N
\frac{\Lambda_i}{ 2 \pi \hbar \Lambda_i (\overline{n}_i(\Lambda_i)
+ \frac{1}{2})}
\end{equation}

The identity relation (\ref{2.64}) can be used to transform to the
following form

\begin{equation}\label{3.4}
z^{-1} = \frac{\omega_0}{ 2 \pi \hbar \lambda_0
(\overline{n}_0(\lambda_0) + \frac{1}{2})} \; \prod_{i=1}^N
\frac{\omega_i}{ 2 \pi \hbar \Lambda_i (\overline{n}_i(\Lambda_i)
+ \frac{1}{2})}
\end{equation}

Putting Eq.(\ref{3.4}) in Eq.(\ref{3.1}) we obtain after
integration over the stable modes

\begin{equation}\label{3.5}
P_{eq} = \frac{\omega_0}{ 2 \pi \hbar \lambda_0
(\overline{n}_0(\lambda_0) + \frac{1}{2})}\;
\frac{\lambda_b}{\omega_b}\;\chi\; \exp \left[ - \;
\frac{\frac{1}{2} \dot \rho^2 + V_1^\ddag - \frac{1}{2}
\lambda_b^2 \rho^2}{\hbar \lambda_0 (\overline{n}_0(\lambda_0) +
\frac{1}{2})} \right]
\end{equation}

where

\begin{equation}\label{3.6}
\chi = \prod_{i=1}^{N}\frac{\hbar \lambda_i
(\overline{n}_i(\lambda_i) + \frac{1}{2})}{\hbar \Lambda_i
(\overline{n}_i(\Lambda_i) + \frac{1}{2})}
\end{equation}

The total energy of the unstable mode is

\begin{equation}\label{3.7}
E = \frac{1}{2} \dot \rho^2 + V_1^\ddag - \frac{1}{2} \lambda_b^2
\rho^2
\end{equation}

The prime quantity for determination of rate constant is the
distribution of energy of the unstable mode. Thus going over to
an energy space so that the co-ordinate $\rho, \;\dot \rho$ are
transformed to $t, E$, respectively with unit Jacobian, the
equilibrium distribution function (\ref{3.5}) is given by,

\begin{equation}\label{3.8}
f_{eq}(E) = \frac{\omega_0}{ 2 \pi \hbar \lambda_0
(\overline{n}_0(\lambda_0) + \frac{1}{2})}\;
\frac{\lambda_b}{\omega_b}\;\chi\; \exp \left[ - \frac{E}{\hbar
\lambda_0 (\overline{n}_0(\lambda_0) + \frac{1}{2})} \right]
\end{equation}

The above distribution is valid for the energy of the unstable
mode $E>V^\ddag$ as well as $E<V^\ddag$.

\subsection{Quantum multidimensional TST rate}

The rate of activated barrier crossing in terms of the equilibrium
probability becomes

\begin{equation}\label{3.9}
\Gamma = \int_{V^\ddag}^\infty f(E)\; dE
\end{equation}

As the unstable mode remains uncoupled from the stable modes the
former mode behaves deterministically and the recrossing does not
occur in this case.

Making use of the distribution (\ref{3.8}) in (\ref{3.9}) we
obtain the rate constant

\begin{equation}\label{3.10}
\Gamma_{QMTST} = \frac{\omega_0}{2
\pi}\;\frac{\lambda_b}{\omega_b}\;\chi\;\exp \left[ -\;
\frac{V^\ddag}{\hbar \lambda_0 (\overline{n}_0(\lambda_0) +
\frac{1}{2})} \right]
\end{equation}

The above expression corresponds to a quantum multidimensional
transition state rate constant. This is central result of this
section. Apart from usual Kramers-Grote-Hynes term
$\lambda_b/\omega_b$ and $\omega_0/2\pi$, the term arising out of
classical transition state result, it contains two important
factors. First, an exponential Arrhenius term where the usual
thermal factor $k_BT$ is replaced by $\hbar \lambda_0
(\overline{n}_0(\lambda_0) + \frac{1}{2})$ includes quantum
effects due to heat bath at very low temperature. In the high
temperature limit it reduces to $k_BT$ and one recovers the usual
Boltzmann factor. This term is essentially an offshoot of a
description of thermal equilibrium by a canonical Wigner
distribution of harmonic oscillator heat bath. Second term $\chi$
can be identified as the quantum correction to Grote-Hynes factor
or more precisely a vacuum corrected generalized Wolynes
contribution for quantum transmission and reflection for the
finite barrier. While usual Wolynes term takes care of the
quantum effects at the higher temperature the factor $\chi$
incorporates quantum effects at arbitrary low temperature. In
what follows we show that the usual Wolynes term and the quantum
rate valid above cross-over temperature can easily be recovered
from $\chi$ in the appropriate limit.

\subsection{Derivation of Wolynes factor from $\chi$ and limiting
rate expression at low temperature above cross-over}

We begin by noting that $\overline{n}(x)$ in $\chi$ which is
given by

\begin{equation}\label{3.11}
\chi = \prod_{i=1}^{N}\frac{\hbar \lambda_i
(\overline{n}_i(\lambda_i) + \frac{1}{2})}{\hbar \Lambda_i
(\overline{n}_i(\Lambda_i) + \frac{1}{2})}
\end{equation}

is the Bose distribution $\overline{n}(x)=(e^{\hbar
x/k_BT}-1)^{-1}$. Neglecting the vacuum contribution $1/2$ from
the terms like $\hbar x(\overline{n}(x)+\frac{1}{2})$ and keeping
only the leading order quantum contribution we obtain

\begin{equation}\label{3.12}
\hbar x\left(\overline{n}(x)+\frac{1}{2}\right) \approx
\frac{\hbar x}{2} \left(\sinh \frac{\hbar x}{2 k_BT}\right)^{-1}
\end{equation}

Therefore $\chi$ reduces to $\Xi$ (say) the Wolynes factor

\begin{equation}\label{3.13}
\Xi = \prod_{i=1}^{N}\frac{\lambda_i \left(\sinh \frac{\hbar
\lambda_i }{2k_BT}\right)^{-1}}{\Lambda_i \left(\sinh \frac{\hbar
\Lambda_i }{2k_BT}\right)^{-1}}
\end{equation}

From identities (\ref{2.63}) and (\ref{2.64}) it follows

\begin{eqnarray}
\prod_{i=1}^{N}\lambda_{i}=\frac{\omega_b}{\lambda_b}\prod_{i=1}^{N}\omega_i\label{3.14}\\
\prod_{i=1}^{N}\Lambda_{i}=\frac{\omega_0}{\lambda_0}\prod_{i=1}^{N}\omega_i\label{3.15}
\end{eqnarray}

respectively, and we have

\begin{equation}\label{3.16}
\prod_{i=1}^{N}\frac{\lambda_i}{\Lambda_{i}}=\frac{\lambda_0
\omega_b}{\omega_0 \lambda_b}
\end{equation}

Making use of the relation (\ref{3.16}) in (\ref{3.13}) we obtain

\begin{equation}\label{3.17}
\Xi = \frac{\lambda_0 \omega_b}{\omega_0 \lambda_b}
\prod_{i=1}^{N}\frac{\sinh ({\hbar \Lambda_i }/{2k_BT})}{\sinh
({\hbar \lambda_i }/{2k_BT})}
\end{equation}

Furthermore $({\lambda_0 \omega_b})/({\omega_0 \lambda_b})$ can
be rewritten as $\frac{\omega_b}{\omega_0}\frac{(\hbar
\lambda_0/k_BT)}{(\hbar \lambda_b/k_BT)}$ which may be
approximated in the form $(\omega_b/\omega_0)\frac{\sinh(\hbar
\lambda_0/2k_BT)}{\sinh(\hbar \lambda_b/2k_BT)}$. Eq.(\ref{3.17})
then reduces to

\begin{equation}\label{3.18}
\Xi = \frac{\omega_b}{\omega_0}\;\frac{\sinh(\hbar
\lambda_0/2k_BT)}{\sinh(\hbar
\lambda_b/2k_BT)}\;\prod_{i=1}^{N}\frac{\sinh ({\hbar \Lambda_i
}/{2k_BT})}{\sinh ({\hbar \lambda_i }/{2k_BT})}
\end{equation}

This is the wellknown Wolynes \cite{woly,wei} expression derived
in eighties as a higher temperature equilibrium quantum
correction to Kramers-Grote-Hynes dynamical factor to Kramers'
rate. Both $\chi$ and the Wolynes factor become unity in the
classical limit. We conclude by noting that unlike Wolynes factor
$\Xi$, $\chi$ is valid below cross-over temperature. The quantity
$\Xi$ approaches unity when the temperature far exceeds the
cross-over temperature. The Wolynes factor may be approximated to
yield a form at low temperature (valid above cross-over
temperature) as
$\exp\left[\frac{\hbar^2}{24}\frac{(\omega_b^2+\omega_0^2)}{(k_B
T )^2}\right]$. This term is responsible for well known $T^2$
enhancement of rate due to quantum effects. With this form of
$\Xi$ and $\exp\left[-\; \frac{V^\ddag}{\hbar \lambda_0
(\overline{n}_0(\lambda_0) + \frac{1}{2})}\right]$ reducing to
$\exp\left[-\; \frac{V^\ddag}{k_B T}\right]$ above the cross-over
regime we obtain from (\ref{3.10}) the rate expression derived
earlier \cite{wei}.

\begin{equation}\label{3.19}
\Gamma=\frac{\omega_0}{2\pi}\;\frac{\lambda_b}{\omega_b}\;\exp
\left[\frac{\hbar^2}{24}\frac{(\omega_b^2+\omega_0^2)}{(k_B T
)^2}\right]\exp\left[-\; \frac{V^\ddag}{k_B T}\right]
\end{equation}

Our general expression (\ref{3.10}) valid in the deep tunneling
regime, \textit{i.e.}, down to vacuum regime.

\section{{\bf Summary and Conclusion}}

Based on a quantum Langevin equation we have constructed a
c-number Hamiltonian for a system plus N-oscillator bath model.
This allows us to formulate a normal mode analysis to realize a
c-number version of the multidimensional transition state theory
and to derive a quantum expression for the total decay rate of
metastable state. The result is valid for arbitrary noise
correlation and temperature down to vacuum limit. The following
pertinent points are noteworthy.

(i) We have shown that the  expression for quantum rate
coefficient is a product of four terms , \textit{e.g.} , classical
well frequency, Kramers'-Grote-Hynes factor, a vacuum corrected
or generalized Wolynes factor representing quantum transmission
and reflection, an exponential term corresponding to Wigner
canonical thermal distribution \textit{i.e.} the generalized
Arrhenius term. Of these the Wigner term and the vacuum corrected
Wolynes term refer to equilibriation of quantum particle in the
well and therefore corresponds to a c-number multidimensional
transition state result. Since the distribution unlike the
Boltzmann is valid even as $T\rightarrow0$, the quantum effect
due to heat bath can be well accounted by these terms even below
the activated tunneling regime.

(ii) The classical limit of the quantum rate expression depends
on Wigner, generalized Wolynes factor. It is easy to see that
they reduce to Arrhenius factor, unity respectively in the limit
$\hbar \omega \ll k_BT$.

(iii) The present theory takes care of activation and tunneling
within a unified description and is equipped to deal with the
rate at temperature down to vacuum limit. This is a distinct
advantage over path integral Monte Carlo method since numerically
the relevant propagator poses serious problem as the temperature
approaches absolute zero.

The quantum theory as presented here is based on a canonical
quantization procedure and a description of equilibrium by
positive definite Wigner's thermal distribution for the bath
rather than path integral or master equation formalism. The
approach has been used and are in use in many related issues
\cite{bk1}. The systematic improvement can be made by taking care
of the quantum corrections of higher orders.

{\bf Acknowledgement}\\
We are thankful to Dr. B. C. Bag for discussions. The authors are
indebted to the Council of Scientific and Industrial Research for
partial financial support under Grant No. 01/(1740)/02/EMR-II.

\end{document}